\documentclass[preprint,showpacs,superscriptaddress,groupedaddress]{aastex}  

\usepackage{graphicx}
\usepackage{dcolumn}
\usepackage{bm}
\usepackage{amssymb}
\usepackage{graphicx}
\usepackage{dcolumn}
\usepackage{bm}
\usepackage{amsmath, amssymb, amsfonts, amsthm}

\newcommand\simlt{\lower.5ex\hbox{$\; \buildrel < \over \sim \;$}}
\newcommand\simgt{\lower.5ex\hbox{$\; \buildrel > \over \sim \;$}}
\newcommand{\gsim}{{\, \lower2truept\hbox{${> \atop\hbox{\raise4truept\hbox{$\sim$}}}$}\,}}

\begin{document}
\tighten
\title{New estimates of the deceleration parameter in weak gravity\footnote{DOI: 10.5281/zenodo.162267}}
\author{Maurice H.P.M. van Putten$^{a}$} 
\affil{$^a$ Room 614, Astronomy and Space Science, Sejong University, 98 Gunja-Dong Gwangin-gu, Seoul 143-747, Korea, email: mvp@sejong.ac.kr}

\begin{abstract}
We consider weak gravity at accelerations $\alpha<a_H$ when Rindler and cosmological horizon collude at $R_H=c/H$, where $c$ is the velocity of light and $H$ is the Hubble parameter. This is manifest in reduced inertia $m$, below the value $m_0$ in Newtonian gravity. Striking evidence for a sharp transition to weak gravity is found in galaxy rotation curves. Their sensitivity to the cosmological background is expressed by correlations to the deceleration
parameter $q=1-(4\pi a_0/cH)^{2}$ and $q=-1/2 -3 (\Omega_b/\sqrt{2}\sqrt{\pi})^{1/2}$, where $a_0$  is Milgrom's scale in the baryonic Tully-Fisher relation of spiral galaxies and $\Omega_b$ is the baryonic matter density. The Planck value $\Omega_b=0.048$ with $H\simeq 73$ km s$^{-1}$ Mpc$^{-1}$ shows $q\simeq-0.85$. Future surveys may determine $Q_0=\left.dq(z)/dz\right|_{z=0}$ to provide a direct test for dynamical dark energy ($Q_0>2.5$) versus $\Lambda$CDM ($Q_0<1$).
\end{abstract}
\maketitle

\section{Introduction}

Dark matter prominently appears in galaxy rotation curves beyond a few kpc of the central bulge of spiral galaxies, in velocity dispersion in galaxy clusters and in cosmological evolution based on analysis of the cosmic micro-wave background (CMB). They have in common dynamics at accelerations below the cosmological scale $a_H=cH_0$, defined by the product of the velocity of light $c$ and the Hubble parameter $H_0$ \citep{fam12}. 

In spiral galaxies, accelerations $\alpha <<a_H$ appears to harden to the baryonic Tully-Fisher inverse distance law \citep{mcg05,mcg11a,mcg11b}, distinct from effectively Newtonian gravitational dynamics within at $\alpha>>a_H$ based on baryonic matter alone. Given ample evidence of dark matter on cosmological scales \citep{pla13}, it is tempting but not imperative to also attribute this apparent anomaly to dark matter concentrations on galactic scales \citep{bek09}. However, the transition acceleration scale 
\begin{eqnarray}
\frac{a_H}{2\pi}\simeq1\mbox{\AA\,s}^{-2}.
\label{EQN_A1}
\end{eqnarray}
is tiny, some four orders below the scales at which Newtonian mechanics and gravitational attraction have been tested. It may reflect a hitherto unappreciated low energy scale in the cosmological vacuum that perturbs motion in possibly unfamiliar ways.

Presently, our Universe is in a state of accelerated expansion that is effectively approaching a de Sitter state driven by a finite density of dark energy \citep{per99}, inferred from a deceleration parameter
\begin{eqnarray}
q=\frac{1}{2}\Omega_M - \Omega_\Lambda < 0
\label{EQN_q}
\end{eqnarray} 
with fractions of cold dark matter $\Omega_M$ and dark energy $\Omega_\Lambda$,, normalized to closure matter density $\rho_c=3H^2_0/8\pi G$ with Newton's constant $G$. The conclusion (\ref{EQN_q}) is based on the three-flat Friedmann-Robertson-Walker (FRW) line-element
\begin{eqnarray}
ds^2=-dt^2 + a(t)^2(dx^2+dy^2+dz^2)
\label{EQN_FRW}
\end{eqnarray}
with dynamical scale factor $a(t)$, $H_0=\left|\dot{a}/a\right|_{z=0}$ and $q=-\ddot{a}a/\dot{a}^2$ evolving according to the theory of general relativity, assuming a no interaction with any low-energy scales in the background vacuum.

To first order, our vacuum is locally Minkowski described by Lorentz invariant light cones $ds=0$ at each point of spacetime. On cosmological scales, the vacuum has a small but finite Gibbons-Hawking temperature  \citep{gib77}
\begin{eqnarray}
T_{dS} = \frac{cH_0}{2\pi k_B}\simeq 3\times 10^{-30}\,\mbox{K}
\label{EQN_TdS}
\end{eqnarray}
associated with the cosmological horizon at $R_H=c/H_0$ in the limit of Lorentz invariant de Sitter space, where $k_B$ denotes the Boltzmann constant and, currently, $H_0 \simeq 73$ km s$^{-1}$Mpc$^{-1}$ \citep{rie16}, at some tension with
$H_0\simeq 67$ km s$^{-1}$Mpc${-1}$ from the CMB \citep{pla13}. 

 \begin{figure}[h]
 	\centerline{\includegraphics[scale=0.30]{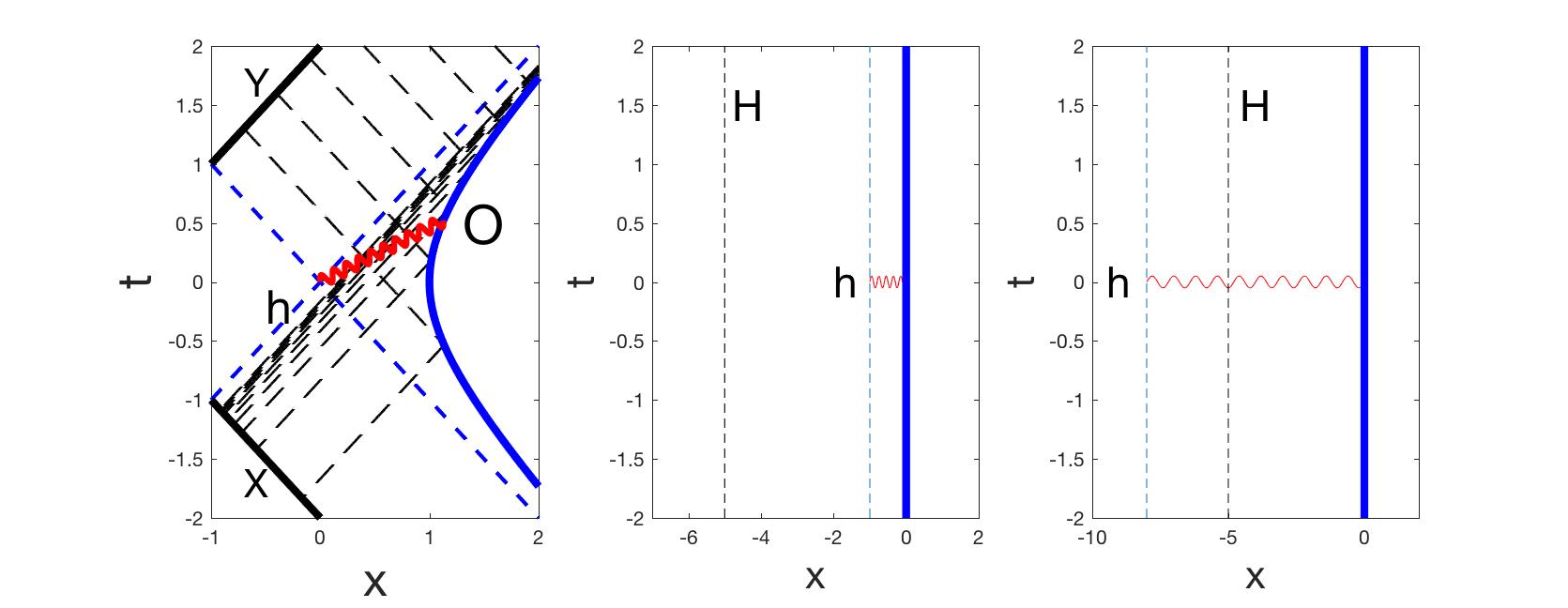}}
 	\caption{(Left panel.) A light cone in Minkowski space forms a horizon surface of a Rindler observer ${\cal O}$ moving at constant acceleration $a$. It assumes a finite temperature of that derives from a map of phase uniform on a null surface element $Y$ in the future into a logarithmic distribution on a null surface element X in the past. Vacuum in the past hereby has a corresponding non-zero state in the future, that appears to ${\cal O}$ at the Unruh temperature (\ref{EQN_A4}). (Middle panel.) A Rindler observer ${\cal O}$ is trailed by an event horizon $h$ at constant distance $\xi$, whose inertia can be attributed to a thermodynamic potential associated with $h$. {\em Strong} and {\em weak} gravity limits (middle and, respectively, right panel) are defined by $\xi$ relative to the distance $R_H$ to the cosmological horizon $H$ as seen by inertial observers. The latter signals a perturbation of ${\cal O}$ inertia below its Newtonian value.}
 	\label{RindlerO}
 \end{figure}
 
In a Newtonian description, stars in galaxies are mutually interacting by gravitational forces that put them on non-inertial trajectories in flat Minkowski spacetime at zero temperature vacuum. Observers along such trajectories see things differently. At acceleration $\alpha$, the world-line of a Rindler observer \citep{bir82} has a time-dependent rapidity $\lambda(\tau) = \alpha\tau$ as a function of eigentime $\tau$. In a Minkowski diagram, the trajectory $x^b(\tau) = \xi\left(\sinh\lambda,\cosh\lambda\right)$ appears curved with $x^b(0)=(0,\xi)$ at a constant distance $\xi=c^2/\alpha$ according to the line-element $ds^2=-dt^2+dx^2$ of 1+1 Minkowski spacetime $(t,x)$. {${\cal O}$ hereby considers itself at a constant distance $\xi$ to an event horizon $h$, given by a light cone with vertex at the origin (Fig. \ref{RindlerO})}, satisfying
\begin{eqnarray}
\alpha\xi=c^2.
\label{EQN_arh}
\end{eqnarray}
Thus, ${\cal O}$'s field of view is limited to a wedge Minkowski spacetime. In this wedge, ${\cal O}$ shares only a finite interval $\Delta\tau = \pi \xi/c$ of its eigentime $\tau$ with inertial observers in Minkowski spacetime.

According to (\ref{EQN_arh}), ${\cal O}$ is trailed by event horizons $h$ at distances $\xi$ inversely proportional to $a$. In unitary holography, a particle of mass $m_0$ at a distance $\xi$ is imaged by information \citep{van15a} 
\begin{eqnarray}
I=2\pi\Delta\varphi
\label{EQN_I}
\end{eqnarray}
on two-dimensional screens, that covers an Einstein area $A_E=8\pi \Delta\varphi\, l_p^2$ in terms of the Compton phase $\Delta \varphi = k\xi$ defined by the wave number $k=mc/\hbar$ and the Planck sized elements of area $l_p^2=G\hbar/c^3$, where $\hbar$ denotes the Planck constant. Holography \citep{bek81} 
is hereby realized by imposing unitarity $P_++P_-=1$ on the probabilities for $m$ to be inside ($P_+$) or outside ($P_-$) a screen of radius $\xi$ about $m_0$. 

 A collusion of the Rindler and cosmological horizon (Fig. \ref{RindlerO}) occurs at when
 $\xi=c^2/\alpha$ satisfies
\begin{eqnarray}
\xi=  R_H.
\label{EQN_A2a}
\end{eqnarray}
In describing the galactic disks of a spiral galaxy of mass $M=M_{11}10^{11}M_0$ with gravitational radius $R_g = GM/c^2$ by Newton's law of gravitational attraction $a_N=GM/r^2$, $a_N\sim a_H=c^2/R_H$, we thus arrive at a transition radius 
 \citep{van16}
 \begin{eqnarray}
 r_t = \sqrt{R_H R_g} = 4.7\,M_{11}^\frac{1}{2}\,\mbox{kpc}
 \label{EQN_A2}
 \end{eqnarray}
 beyond which we reach weak gravity
 \begin{eqnarray}
 \xi > R_H.
 \label{EQN_W1}
 \end{eqnarray}
 as the Rindler horizon $h$ drops beyond $H$.
 
 This expression (\ref{EQN_A2}) serves as an order of magnitude estimate. Implications for gravitational attraction requires a detailed consideration on the properties of $h$ and $H$, since $h$ is defined for ${\cal O}$ while the cosmological horizon is ab initio defined for inertial observers. Nevertheless, (\ref{EQN_A2}) suffices to identify a transition between strong and weak gravity, from Newtonian to non-Newtonian behavior in galaxy rotation curves.

 In the following, we pursue (\ref{EQN_A2}) further in weak gravity accompanying baryonic matter. To start, we identify a holographic origin of inertia associated thermodynamic properties of the Rindler event horizon. Rest mass hereby equals the potential energy in the gravitational field associated with the Rindler horizon. In Minkowski space, inertial and total mass-energy are hereby identical. The same holds true in spherical symmetry about the event horizons of black holes in asymptotically flat spacetimes. We extend this approach to the homogeneous and isotropic de Sitter and, more generally, FRW cosmologies. Our approach identifies a breakdown of the equivalence principle at non-relativistic vacuum temperatures. It results in a relation between total mass energy and baryonic mass-energy, based on dispersive behavior in cosmological holography at low vacuum temperatures on the order of (\ref{EQN_TdS}).
 
\section{Holographic inertia of Rindler observers}
 
 In Minkowski spacetime, Rindler observers detect the Unruh temperature
  \citep{unr76}
 \begin{eqnarray}
 k_BT_U = \frac{\hbar \alpha}{2\pi c}
 \label{EQN_A4}
 \end{eqnarray}
 associated with the surface gravity of $h$. 
 To highlight its basic premises in common with (\ref{EQN_TdS}), we recall that (\ref{EQN_A4}) derives from a Bogoliubov map, by ray tracing between radiative Hilbert spaces ${\cal H}_X$ and ${\cal H}_Y$ in the past and, respectively, future \citep{bir82}, here shown in Fig. \ref{RindlerO} 
 with null surface elements $X$ and $Y$ as indicated. 
 A choice of uniform distribution of phase on $Y$ defines a given positive frequency state of ${\cal H}_Y$ relative to $\left|\right. 0 \left.\right>$. The ray tracing map thus introduces a Fourier transform on $X$ of a logarithmic distribution of phase $\phi(x)=\alpha \ln x$ over $X=(0,\infty)$,
 giving rise to a thermal spectrum $\left|\beta\right|^2  = (e^{2\pi \alpha} -1)^{-1}$ \citep{bir82}.
  
 In this process, a Rindler observer, according to its eigentime, detects a wave front with exponential divergence in frequency into the future. Starting from the true vacuum state of $X$, the inverse of our map gives a thermal distribution of positive energy photon number (per unit time per unit frequency) $\left|\beta\right|^2$ out to the future through $Y$ \citep[][]{bir82}. 

 At a distance $\xi$ to the event horizon $h$ of a Rindler observer, changes in $\xi$ introduce a corresponding change in entropy $dS=-k_BdI$ at $T_U$ of $\alpha$ on $h$, and $T_U=k_B^{-1}\left({\partial S}/{\partial m_0c^2}\right)^{-1} \equiv {a\hbar}({2\pi c})$
is also a thermodynamic temperature. Inertia of Rindler observers hereby satisfies
\begin{eqnarray}
F=-T_U \frac{dS}{d\xi} = \left(\frac{\hbar \alpha}{2\pi c}\right) 
\left( 2\pi\frac{m c}{\hbar}\right) = m_0\alpha
\label{EQN_A5}
\end{eqnarray}
with associated potential energy relative to $h$ equal to its rest mass energy,
\begin{eqnarray}
U = \int_h^{\cal O} F ds = F\xi = m_0c^2,
\label{EQN_U1}
\end{eqnarray}
by virtue of $\alpha$ being constant along $\xi$. 

In (\ref{EQN_U1}), inertia $m$ originates in the thermodynamic potential $U$ associated with the Rindler horizon $h$. This implies that $m$ will be perturbed away from $m_0$ in (\ref{EQN_U1}) whenever $h$ drops beyond the cosmological horizon $H$. The quantify this, we next consider the geometry and thermodynamic properties of the latter.

\section{Intrinsic and extrinsic surface gravity}

In FRW cosmologies, cosmological horizons are trapped surfaces defined in hypersurfaces of constant cosmic time $t$. Given a unit normal $s^i$ and an extrensic curvature $K_{ij}=\nabla_is_j$, they satisfy $\nabla_is^i + s_is_iK^{ij}-K=0$ \citep{coo00}. 
In three-flat FRW cosmologies with Hubble parameter $H$, this reduces to $\nabla_is^i=2H$, i.e., $R_H=c/H$ as stated before. A covariant derivation of surface gravity obtains from geodesic deviation between a pair of generators in four-dimensionlal spacetime $(t,r,\theta,\varphi)$ centered about an inertial observer. A tangent $k^b$ to such generator is null, $k^tk^t-a^2r^2k^\theta k^\theta=0$. A spacelike separation $l$ in $\Sigma_t$ between such null-generators, initially along the unit vector $m^c=a^{-1}r^{-1}(0,0,1,0)$ experiences a change in length with tip-to-tip acceleration, given by the equation of geodesic deviation along $m^c$. With cosmic time $t$ as an affine parameter, $k^t=1$, the acceleration $\alpha^b$ of one tip satisfies
\begin{eqnarray}
\alpha^b = - \frac{1}{2}R_{aec}^{~~~b}k^ak^cm^e = - a\frac{1}{2}\left( 1 - q \right) H,
\label{EQN_alpha1}
\end{eqnarray}  
where $R_{abcd}$ denotes the Riemann tensor of the FRW line-element \citep{pir57}. 
In magnitude, $\alpha = \sqrt{\alpha^c\alpha_c}$ satisfies the {\em intrinsic} surface gravity
\begin{eqnarray}
a_H = \frac{1}{2}\left( 1 - q \right) H.
\label{EQN_alpha2}
\end{eqnarray}  
Once again, $T_H = \hbar a_H/2\pi c$ is an associated thermodyamic temperature \citep{cai05}.

The same arguments apply to black hole event horizons, upon considering a null-tangent $k^b = (k^t,k^\theta)$ of the event horizon of a Schwarzschild black hole with mass $M$. It has the intrinsic surface gravity
\begin{eqnarray}
a_H = \frac{GM}{r^2_H}=\frac{c^4}{4GM}
\end{eqnarray}
equal to the {\em extrinsic} surface gravity
\begin{eqnarray}
a_H^\prime \equiv - \lim_{r\rightarrow r_H} \frac{dA}{ds}
\end{eqnarray}
in the Schwarzschild line-element with $ds=dr/A$ and redshift factor $A(r)=\sqrt{1-2R_g/r}$, $R_g=GM/c^2$. This suface gravity is conform Hawking's thermodynamic temperature \citep{haw75}
$T_H = {\hbar a_H}({2\pi})$ akin to (\ref{EQN_U1}): a test particle of rest mass $m<<M$ taken from $H$ to infinity is associated with entropic work $W=T_H \delta S = m T_H \left( {\partial S_H}/{\partial Mc^2}\right) = mc^2$, based on the Bekenstein-Hawking entropy $S_H = (1/4) k_BA_H/l_p^2$ \citep{bek73,haw75}, where $l_p=\sqrt{G\hbar/c^3}$ denotes the Planck length associated with Planck's constant $\hbar$. In the Schwarzschild line-element of black holes, the same result obtains from the line-integral of force $F=mGM/r^2$ over distance $s$ from $R_H=2R_g$ to infinfity, i.e., $U(r)=\int_{R_H}^r Fds=\sqrt{1-2M/r}\,mc^2$. This recovers
\begin{eqnarray}
U=\int_{R_H}^\infty Fds=mc^2
\label{EQN_U2}
\end{eqnarray}
as the gravitational binding energy of a mass element $m$ in the black hole.
Hidden behind the event horizon, the source of the gravitational field is again unseen \citep{pen65}. In unitary holography, the black hole and cosmological event horizons define the minimum, respectively, maximum sized holographic screens with $A_E=A$.

\section{A holographic origin of dark energy}

In a Cartesian coordinate system, $dy=rd\theta$ at $r=R_H$, the geodesic deviation equation (\ref{EQN_alpha1}) gives $\ddot{l}\delta_{3d} = - lR_{abcd}k^ak^cm^b = - a\left( 1 - q \right) H^2 l$ for the acceleration of separation $l$ parallelly transported between two null-generators of a horizon null-surface. Upon taking the norm of $\delta_{3d}$, this defines an harmonic oscillation
\begin{eqnarray}
\ddot{u} + \omega^2_0 u =0,~~\omega_0 = \sqrt{1-q}H_0 
\label{EQN_alpha5}
\end{eqnarray}
that, in a holographic approach, extends to a wave equation in spacetime within,
\begin{eqnarray}
\Box u + \Lambda u =0,~~\omega =\sqrt{c^2k^2+ \hbar^2\Lambda},
\label{EQN_alpha6}
\end{eqnarray}
where we \citep{van15b}
\begin{eqnarray}
\Lambda = (1-q)H^2.
\label{EQN_LAMBDA}
\end{eqnarray} 
The wave equation (\ref{EQN_alpha6}) will be recognized to describe electromagnetic and linearized gravitational waves in de Sitter space, considered in a Lorenz gauge on the U(1) and, respectively, SO(3,1) connection \citep{wal84,van96}, by coupling to the Ricci tensor $R_{ab}=c^{-2}\Lambda g_{ab}$. Importantly, (\ref{EQN_alpha6}) reveals dispersion at low energies.

As illustrated by Rindler observers in Minkowski spacetime, the equivalence principle describes sameness of Lorentz factors in Minkowski space and potentials in static gravitational fields. For an inertial observer in Minkowski spacetime, a slice of constant time represents an equipotential hypersurface, while a Rindler observer identifies a potential gradient in hypersurfaces of constant eigentime. Their equivalence satisfies conservation of total energy, in that the observable effects of a Lorentz factor in the first and a potential drop in the second are equivalent. In (\ref{EQN_U1}) and (\ref{EQN_U2}), the equivalence principle applies independent of geometry, in uniform and spherically symmetric gravitational fields alike. In de Sitter cosmologies, surfaces of constant cosmic time are equipotential hypersurfaces as in Minkowski space, endowed with geodesic expansion described by the Hubble parameter with associated Lorentz factor $\Gamma = 1/\sqrt{1-H^2r^2}$ as a function of distance $r$ to an inertial observer. 

According to aforementioned equivalence principle, spacetime about an inertial observer is equivalently static and spherically symmetric with a gravitational potential $U(r)=1/\Gamma$, at least in the case of static event horizons. The equipotential surface of constant cosmic time $t$ in (\ref{EQN_FRW}) now becomes equivalent to surfaces on which test particles satisfy $U\Gamma = 1$. The net result is the well-known static de Sitter universe with an extension to a central mass $m$ described by
\begin{eqnarray}
U(r)=\sqrt{1-\frac{2M}{r} - \frac{1}{3}\Lambda r^2}.
\label{EQN_U3}
\end{eqnarray}
Imposing $U(R_H)=0$ in this equivalence of a static, spherically universe with (\ref{EQN_LAMBDA}), we obtain 
\begin{eqnarray}
\Omega_M + \Omega_\Lambda = 1.
\label{EQN_U3b}
\end{eqnarray}
with \citep{van15b}
\begin{eqnarray}
\Omega_M= \frac{1}{3}(2+q),~~\Omega_\Lambda = \frac{1}{3}(1-q)
\label{EQN_M}
\end{eqnarray}
in terms of the mass density $M/M_0$ at a total mass-energy $M_0 = \rho_c V=R_0/2$ within the volume $V=(4\pi/3)R_H^3$. With $ds=dr/U$, we note the extrinsic surface gravity at $U(R_H)=0$ 
\begin{eqnarray}
a_H^\prime =  - \lim_{r\rightarrow R_H}\frac{dU}{ds} = \left[ 2\Omega_\Lambda - \Omega_M \right] cH = -qcH.
\label{EQN_alpha2b}
\end{eqnarray}
Thus, $a_H^\prime$ is directly related to the Hubble flow through the cosmological horizon, positive for ingoing ($q>0$) and negative for outgoing ($q<0$) flows. Outgoing Hubble flows conform to the Newtonian regime of negative gravitational binding energy.

In de Sitter spacetime $(q=-1)$, $a_H^\prime$ equals $a_H$ in (\ref{EQN_alpha2}). By virtue of constant radius, its horizon is akin to Rindler horizons in Minkowski spacetime. In identifying the equivalence principle with $a_H^\prime=a_H$ leading to (\ref{EQN_U1}) and (\ref{EQN_U2}). The mass-energy $mc^2$ of a test particle represents the gravitational binding energy in (\ref{EQN_U3}) in taking it out from the horizon at $R_H=c/H$.
{\em The discrepancy between intrinsic and extrinsic surfrace gravity,
\begin{eqnarray}
a_H -  a_H^\prime = \left(\frac{1+q}{2}\right)cH
\label{EQN_Da}
\end{eqnarray}
quantifies breakdown of the equivalence principle in non-de Sitter cosmologies.}
Likewise, screen temperature $\hbar a_H/2\pi c$ is generally not equal to local vacuum temperature $\hbar\left|a_H^\prime\right|/2\pi c$. According to (\ref{EQN_alpha2}) and (\ref{EQN_alpha6}), the association of intrinsic surface gravity with horizon temperature and dark energy defines $q\le 1$ as a positive energy condition. 

\section{Perturbed inertia in galaxy rotation curves at weak gravity}

In weak gravity (\ref{EQN_W1}), (\ref{EQN_U1}) is reduced by integration limited to $R_H$, whereby $U=mc^2$ satisfies $mc^2\propto \left({R_H}/{\xi}\right) m_0c^2$. As a holographic condition, we impose equality in the number of screen modes and radial bulk modes,
\begin{eqnarray}
\frac{mc^2}{\frac{1}{2}\sqrt{\hbar^2\Lambda + c^2p^2}} = \frac{(R_H/\xi)m_0c^2}{k_B\sqrt{T_H^2+c^2p^2}}
\label{EQN_B1}
\end{eqnarray}
where $p$ refers to a momentum coupling between screen and bulk modes and $T_H$ is the temperature of the cosmological horizon associated with intrinsic curvature (\ref{EQN_alpha2b})
\begin{eqnarray}
T_H = \left(\frac{1-q}{2}\right)T_{dS}.
\label{EQN_TH}
\end{eqnarray}
With $R_H/\xi = \alpha/a_H$, it follows that $m=m_0\alpha/(2Ba_H)$ with
\begin{eqnarray}
B(p) = \frac{k_B\sqrt{T^2_H+c^2p^2}}{\sqrt{\hbar^2\Lambda+c^2p^2}}.
\label{EQN_Bp}
\end{eqnarray}
Given a gravitational attraction $F_N = GMm_0/r^2 \equiv a_Nm_0$ by a baryonic mass $M$, we thus arrive at an acceleration 
\begin{eqnarray}
\alpha = \sqrt{2Ba_Ha_N}. 
\label{EQN_B0}
\end{eqnarray}

At very low momenta corresponding to
\begin{eqnarray}
\xi >> R_H
\label{EQN_W2}
\end{eqnarray}
$B\simeq B(0)=k_BT_H/\hbar\sqrt{\Lambda}=\sqrt{(1-q)/2}/(2\sqrt{2}\pi)$ in (\ref{EQN_B0}) with (\ref{EQN_LAMBDA}) captures the Tully-Fisher law equivalent to Milgrom's inverse distance law $\alpha = \sqrt{a_0a_N}$ of gravity with Milgrom's constant $a_0$ \citep{mil83}, here $a_0=2Ba_H$ with \citep{van16}
\begin{eqnarray}
a_0 = \frac{cH}{\sqrt{2}\pi}\left({\frac{1-q}{2}}\right)^\frac{1}{2}.
\label{EQN_a0}
\end{eqnarray}
Presently, we have
\begin{eqnarray}
\left.a_0\right|_{z=0} =  1.2\,h_0\,\left(\frac{1-q}{1.8}\right)^\frac{1}{2}\,\mbox{\AA}\,\mbox{s}^{-2}.
\label{EQN_a1}
\end{eqnarray}
where $h_0$ refers to the present Hubble parameter normalized to $H_0\simeq 70\,\mbox{km s}^{-1}\mbox{Mpc}^{-1}$. The weak gravity range (\ref{EQN_W2}) of (\ref{EQN_B0}) is well known for its description of galaxy rotation curves. However, a constant Milgrom parameter $a_0$, does not permit a continuous transition to strong gavity in the sense of $\xi > R_H$. This observation has led to various phenomenological interpolation functions for $a_0$ as a function of acceleration \citep{fam12}.

\begin{figure}[h]
	\centerline{\includegraphics[scale=0.35]{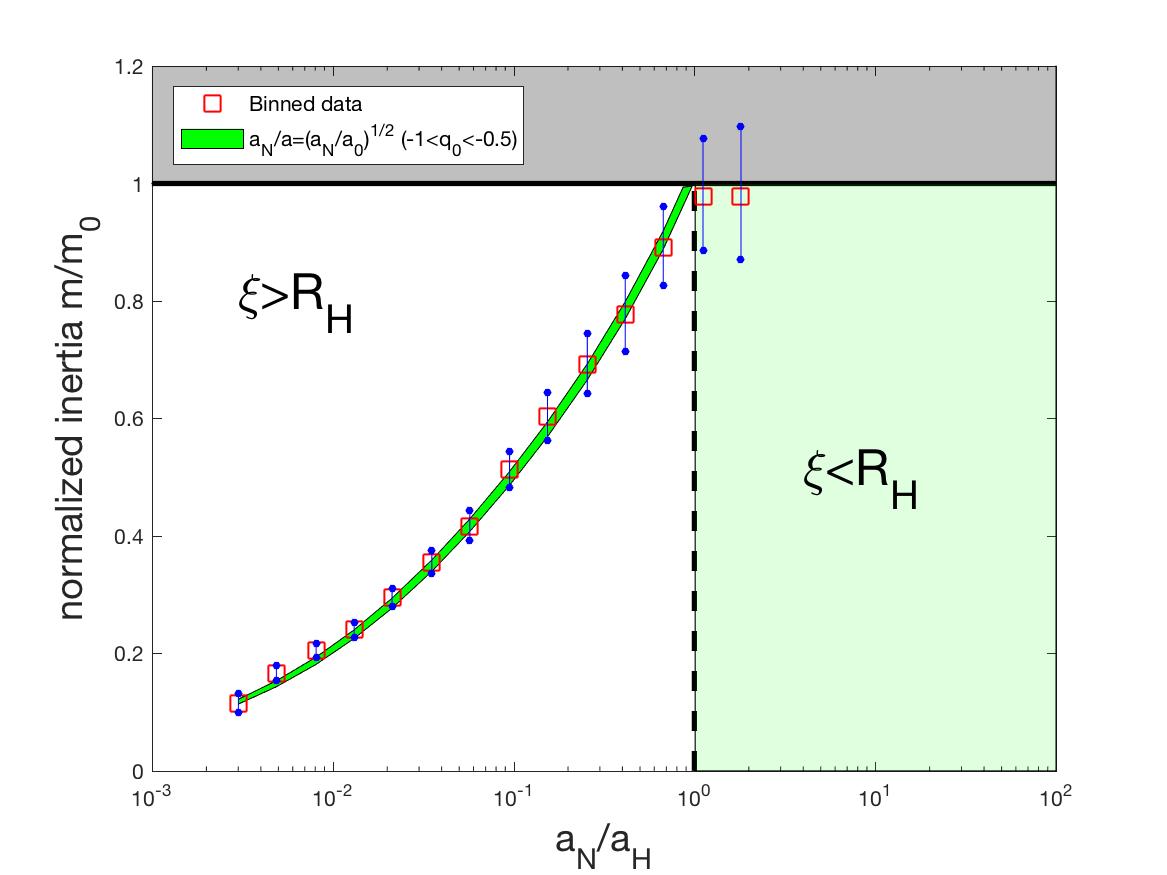}}
	\caption{Accelerations $\alpha=a_H$ define a transition to weak gravity (\ref{EQN_W1}) as Rindler and cosmological horizon collude. For $\alpha<a_H$, inertia $m$ drops below the Newtonian value $m_0$. This transition is sharp as manifest in galaxy rotation curves, here plotted as $m/m_0$ as a function of Newtonian gravitational acceleration $a_N/a_H$ based on baryonic matter content. Shown are binned data accompanied by $3\sigma$ uncertainties. The curved green band is the theoretical curve covering $-1<q_0<-0.5$ with $H_0=73$ km s$^{-1}$Mpc$^{-1}$. (Data from \cite{fam12b}).}
	\label{Fig:rotation}
\end{figure}

In the transition region between (\ref{EQN_W2}) and (\ref{EQN_W1}), we encounter the momentum dependent coefficient $B=B(p)$. In associating $p=\left|{\bf p}\right|$ with an isotropic thermal distribution of vacuum fluctuations, we propose a thermal average 
\begin{eqnarray}
\left< B \right>_T = \frac{1}{W} \int_0^\infty B(p) e^{-\frac{E}{k_BT}} \left(4\pi p^2 \right)dp,~~W=\int_0^\infty e^{-\frac{E}{k_BT}} \left(4\pi p^2\right) dp
\label{EQN_B}
\end{eqnarray}
with $E=\sqrt{\hbar^2\Lambda+c^2p^2}-\hbar\sqrt{\Lambda}$. For numerical evaluation, we express (\ref{EQN_B}) as 
\begin{eqnarray}
\left< B \right>_y = \frac{1}{W} \int_0^\infty f(x) e^{-e/y} x^2 dx,~~W=\int_0^\infty e^{-e/y} x^2dx
\label{EQN_Bx}
\end{eqnarray}
with $f(x)=\sqrt{1+x^2}/\sqrt{A+x^2}$, $e=\sqrt{A+x^2}-A^{1/2}$, $A=16\pi^2/(1-q)$. 
With $y={a_N}/{a_H}$, the thermal average $\left<B\right>_y$ has the property 
that $\left<B\right>_1 \simeq 1/2$ (to be precise, $\left<B\right>_1=0.4691$ for $q=-0.5$ and $\left<B\right>_1= 0.4995$ for $q=-1$), effectively ensuring continuity $\alpha\simeq a_H$ in the transition to weak gravity (\ref{EQN_W1}), while
$\left<B\right>_y\rightarrow B(0)$ as $y\rightarrow0$.

Fig. 2 shows a confrontation of the perturbed inertia as a function of Newtonian acceleration $a_N$ based on the observed baryonic matter in spiral galaxies. 

The data in Fig. 2 show a remarkably sharp transition at $\alpha=a_H$, here identified with (\ref{EQN_W1}). With (\ref{EQN_B}), (\ref{EQN_B0}) gives a well-defined and accurate transition to Milgrom's law at $\alpha<<a_H$.

\section{Detecting deceleration}

Measurements of Hubble parameter $H=H(z)$ and the deceleration parameter $q(z)$ provide a powerful probe of cosmic expansion that may be obtained from local surveys independent of cosmological models.

Inverting (\ref{EQN_a0}) gives measurement of the deceleration parameter,
\begin{eqnarray}
q(z) = 1 - \left( \frac{4\pi a_0(z)}{cH(z)}\right)^{2}.
\label{EQN_qr}
\end{eqnarray}
For $q_0=q(0)$, we encounter uncertainties in $a_0(0)$ and $H_0=H(0)$. A best fit to the data of Fig. 2 based on least square errors shows $q_0 = -0.6$ based on the data points in the weak gravity range (\ref{EQN_W1}) and $q_0 = -0.8$ when augmented with the first data point about the transition to strong gravity. From local surveys, a typical  Hubble parameter is $H_0=73$ km s$^{-1}$Mpc$^{-1}$ in tension by about 7\% with $H_0\simeq 68$ km s$^{-1}$Mpc$^{-1}$ obtained from CMB analysis. Based on (\ref{EQN_qr}), uncertainties in $H$ contribute to $\delta q\simeq 0.05-0.10$. Since $\delta q/q = -2\delta H/H$ in (\ref{EQN_qr}), (\ref{EQN_qr}) gives a mild constraint $-0.9<q_0<-0.5$ on $q_0$.

Alternatively, we may turn to data from the CMB. The dark matter density $\Omega_M$ in (\ref{EQN_U3b}) is similar but not identical to a cold dark matter density $\Omega_m$ in $\Lambda$CDM. In particular, $\Omega_M=5/6$ and $\Omega_m=1$ in the matter dominated era $q=1/2$. This precludes a direct comparison of $\Omega_M$ with dark matter measurements obtained by fitting $\Lambda$CDM to Planck data. In fact, their somewhat different evolution, here attributed to a dynamical (\ref{EQN_LAMBDA}) versus constant dark energy density, is conceivably part of the notorious tension in, e.g., the Hubble parameter in $\Lambda$CDM extracted from the CMB or from the Local Universe \citep{rie16}. Planck estimates of $\Omega_\Lambda$ and the baryonic matter content $\Omega_b$ are likewise somewhat model dependent. Since $\Lambda$CDM provides a leading order approximation to better than 10\%, we next consider the following.

For $r=R_H$, (\ref{EQN_Da}) and (\ref{EQN_a1}) define $E=a_HR_H^2$ and $E_m=a_H^\prime R_H^2$, satisfying $E = (1+q)M_0 + \Omega_m$ with matter density $\Omega_{m} = 2M_0\sqrt{\beta_0\Omega_b}$, where $M_0=R_H/2$ denotes the total mass energy $\rho_c(4\pi/3)R_H^2$. Let $\Omega_M = E/M_0$. Then $\Omega_M = 1+q + \sqrt{{2\Omega_b}/{\sqrt{\pi}}}$. Consistency with (\ref{EQN_M}) obtains when $({1}/{3})(1+2q) + \sqrt{{2\Omega_b}/{\sqrt{\pi}}} = 0$, that is
\begin{eqnarray}
q=-\frac{1}{2} - 3 \sqrt{\frac{\Omega_b}{2\sqrt{\pi}}}.
\label{EQN_qc}
\end{eqnarray}
The Planck value $\Omega_b = 0.048$ gives $\Omega_m = 0.2327$ and a deceleration parameter
\begin{eqnarray}
q_0= - 0.85.
\label{EQN_qb}
\end{eqnarray}
Since $\delta q = -(1/2)\delta \Omega_b/\Omega_b$ in (\ref{EQN_qc}), (\ref{EQN_qc}) gives a relatively secure constraint on $q_0$, more so than (\ref{EQN_qr}) for the reasons stated above.
 
\section{Conclusions and outlook}

High resolution galaxy rotation curve data \citep{fam12b} reveal a remarkably sharp 
transition in inertia (Fig. 2) at the collusion of the Rindler and cosmological horizon (\ref{EQN_A2a}). In practical terms, it defines a transition to an inverse distance law of gravity in (\ref{EQN_W1}), the asymptotic regime (\ref{EQN_W2}) of which is known as the Tully-Fisher law or, equivalently, Milgrom's law (\ref{EQN_B0}) with (\ref{EQN_a0}). The transition of (\ref{EQN_A2a}) between (\ref{EQN_A2a}) and the latter is described by a running $B(p)$ value, modeled by a thermal average $<B>_y$ that takes $2<B>_Ta_H$ from effectively $a_H$ down to $a_0$ as a function of $y=a_N/a_H$. The outcome agrees with the galaxy rotation curves within measurement uncertainties.

In identifying inertia with a thermodynamic potential that is perturbed in weak gravity (\ref{EQN_A1}), galaxy rotation curves are sensitive to the cosmological background. In particular, sensitivity to $q$ in (\ref{EQN_a0}) gives rise to (\ref{EQN_qr}), that may be probed for redshift dependence in future galaxy rotation surveys to determine $Q(z) = {dq(z)}/{dz}$,
\begin{eqnarray}
Q(z)=2(1-q) \left( H^{-1}\frac{dH}{dz} - a_0^{-1}\frac{da_0(z)}{dz}\right)
\label{EQN_Qz1}
\end{eqnarray} 
with $H^{-1}dH/dz=(1+q)/(1+z)$. If sufficiently well resolved, measurement of $Q_0=Q(0)$ in (\ref{EQN_Qz1}) suggests a direct test of dynamical versus static dark energy, i.e., (\ref{EQN_LAMBDA}) versus  $\Lambda$CDM, with \citep{van16}
\begin{eqnarray}
Q_{0,dyn} >2.5,~~Q_{0,stat}<1.
\label{EQN_Qz2}
\end{eqnarray}

In our model of weak gravity, the mass scale $\hbar\sqrt{\Lambda}\simeq 2\sqrt{2}\pi k_BT_{dS}=2\times 10^{-33}$ eV sets a minimum to that of dark matter. While pairing in condensations may conceivably increase its mass \cite[e.g.][]{van10}, this low mass is expected to effectively limit clustering to the scale of galaxy clusters \citep[e.g.][]{vik15}. As such, it will be undetectable by laboratory experiments seeking direct interactions with baryonic matter. Instead, it may be probed by (\ref{EQN_qb}) and (\ref{EQN_qr}), and the associated dark energy by (\ref{EQN_Qz1}-\ref{EQN_Qz2}). Alternatively, satellite based, free fall Cavendish type experiments may probe (\ref{EQN_A2}), rescaled to laboratory size and mass $M$ to $r_t = \sqrt{R_H R_g} =  \left(M/1\mbox{g}\right)^\frac{1}{2}\,\mbox{cm}$.
 
{\bf Acknowledgement.} This report is supported in part by the Korean National Research Foundation under Grants 2015R1D1A1A01059793 and 2016R1A5A1013277.

\end{document}